\documentclass[prx,twocolumn,superscriptaddress,longbibliography]{revtex4-2}
\usepackage{bm}
\usepackage{amsmath, amsfonts}
\usepackage{amssymb}
\usepackage{times}
\usepackage{graphicx}
\usepackage{xcolor}
\usepackage{mathtools}
\usepackage{physics}
\usepackage{mleftright}
\usepackage{float}
\usepackage{setspace}

\usepackage[colorlinks=true,allcolors=blue]{hyperref}
\usepackage[capitalise,nameinlink]{cleveref}
\usepackage{tikz}
\crefname{section}{Sec.}{Figs.}
\graphicspath{{figures/}}
\begin{document}

\title{Observation of coherent flux-charge interaction in a gate-tunable fluxonium}

\author{Brian D. Isakov}
\thanks{These authors contributed equally to this work.}
\affiliation{Department of Electrical, Computer \& Energy Engineering, University of Colorado Boulder, Boulder, CO 80309, USA}

\author{Shikhar Singh}
\thanks{These authors contributed equally to this work.}
\affiliation{NNF Quantum Computing Programme, Niels Bohr Institute, University of Copenhagen, 2100 Copenhagen, Denmark}

\author{Adrian Parra-Rodriguez}
\affiliation{Technical University of Munich, TUM School of Natural Sciences, Physics Department, 85748 Garching, Germany}
\affiliation{Walther-Meißner-Institut, Bayerische Akademie der Wissenschaften, 85748 Garching, Germany}
\affiliation{Munich Center for Quantum Science and Technology (MCQST), 80799 Munich, Germany}

\author{David Feldstein-Bofill}
\affiliation{NNF Quantum Computing Programme, Niels Bohr Institute, University of Copenhagen, 2100 Copenhagen, Denmark}

\author{Zhenhai Sun}
\affiliation{NNF Quantum Computing Programme, Niels Bohr Institute, University of Copenhagen, 2100 Copenhagen, Denmark}

\author{Anders Kringhøj}
\affiliation{NNF Quantum Computing Programme, Niels Bohr Institute, University of Copenhagen, 2100 Copenhagen, Denmark}

\author{Svend Krøjer}
\affiliation{NNF Quantum Computing Programme, Niels Bohr Institute, University of Copenhagen, 2100 Copenhagen, Denmark}

\author{Alexandre Blais}
\address{Institut quantique \& D\'epartement de Physique, Universit\'e de Sherbrooke, Sherbrooke J1K2R1, Qu\'ebec, Canada}

\author{Morten Kjaergaard}
\thanks{Corresponding author}
\email{mkjaergaard@nbi.ku.dk}
\affiliation{NNF Quantum Computing Programme, Niels Bohr Institute, University of Copenhagen, 2100 Copenhagen, Denmark}

\author{Andr\'as Gyenis}
\thanks{Corresponding author}
\email{andras.gyenis@colorado.edu}
\affiliation{Department of Electrical, Computer \& Energy Engineering, University of Colorado Boulder, Boulder, CO 80309, USA}
\affiliation{Department of Physics, University of Colorado Boulder, Boulder, CO 80309, USA}

\begin{abstract}
    Interactions that mix conjugate variables, such as the flux through a circuit element and the charge across it, lie outside the reach of the elementary couplings of superconducting circuits. Capacitors connect charge to charge, and inductors connect flux to flux, while no two-terminal element couples flux to charge directly. A native flux-charge coupling would thus serve as a circuit primitive in its own right, opening direct routes to non-reciprocity, protected modes, and unconventional readout. In this work, we demonstrate a flux-charge coupling by harnessing a voltage-tunable Josephson junction with parametrically modulated critical current, which mediates the interaction between a classical charge variable and a quantum flux operator. Relying on parity-selection rules in a hybrid superconducting-semiconductor fluxonium, we isolate the flux-charge coupling from other parasitic capacitive contributions and perform cross-quadrature-activated coherent control of states. Critically, we realize a flux-charge coupling that scales linearly with driving amplitude while keeping the transition energy first-order-insensitive to gate voltage. Such unconventional interaction broadens the toolbox of superconducting circuits with a critical missing component that enables the coherent coupling of conjugate variables.
\end{abstract}

\maketitle

\subsection*{Introduction}

Quantum theories hinge on the commutation relations between conjugate operator pairs, such as the position $\hat{x}$ and momentum $\hat{p}$ operators. In elementary Hamiltonians, these operators appear separately, in the kinetic $K(\hat{p})$ and potential energy $V(\hat{x})$ of a particle. Moving beyond these simple cases is beneficial because linking operators of opposite quadratures can lead to a rich set of quantum phenomena in which reciprocity is broken~\cite{10.21468/SciPostPhysLectNotes.44, Potton_2004, Fruchart2021, PRXQuantum.3.020201, Wanjura2023}. For instance, cross-quadrature coupling emerges in systems where the angular momentum plays an essential role due to magnetic fields or temporal modulations of a drive, leading to topological one-way edge states, metamaterials, and gyration~\cite{PhysRevLett.45.494, Roushan2017, Rechtsman2013,Carusotto2020,7mfg-5h1x}. Similarly, squeezing---implemented with driven nonlinearities---is governed by a Hamiltonian that couples the two conjugate quadratures directly, $\hat H_{\rm sq}\propto i(\hat a^{\dagger 2}-\hat a^2)
=(\hat x\hat p+\hat p\hat x)$, where $\hat{a}$ and $\hat{a}^\dagger$ are the standard creation and annihilation operators. The squeezing produced by such cross-quadrature interactions is in turn a resource for applications ranging from quantum sensing to the preparation of protected and error-correctable states~\cite{PhysRevLett.55.2409,PhysRevLett.59.2153,Grimm2020,PhysRevA.64.012310}.

\begin{figure*}
\centering
\includegraphics[width = \textwidth]{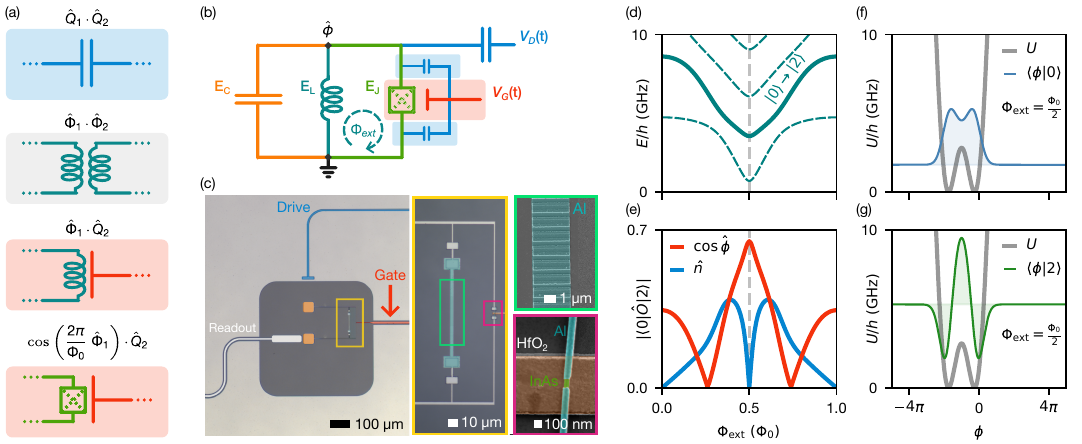}
\caption{
\label{fig:fig1} 
\textbf{Flux-charge coupling in a nanowire fluxonium}. a, Different coupling methods in superconducting circuits. Capacitive or inductive interactions between two modes lead to charge-charge or flux-flux coupling operators, while a composite flux-charge interaction can be accessed by a hypothetical capacitor-inductor element. A gate-tunable Josephson junction with parametrically modulated gate voltage can provide such an interaction in practice. b, Circuit diagram of the superconductor-semiconductor hybrid nanowire fluxonium, where a drive line is capacitively coupled (blue), and a gate line placed underneath the junction provides the flux-charge coupling (red) in addition to parasitic capacitive coupling (blue). c, False-colored optical and scanning electron microscope images of the gate-tunable fluxonium. The inductor is formed using an Al/AlO$_x$ Josephson junction array, and the semiconductor junction is formed with an etched InAs nanowire with epitaxially grown Al on all facets. d, e, Numerical simulation of fluxonium transitions and matrix elements of the $|0\rangle\rightarrow|2\rangle$ transition as a function of external flux with $E_J/h = 3.28\,$GHz, $E_L/h = 1.00\,$GHz, and $E_C/h = 0.88\,$GHz. Due to parity selection rules, the flux-charge coupling can be isolated at $\Phi_\mathrm{ext}=\Phi_0/2$. f, g, Fluxonium potential and wavefunctions for the ground and second excited states at $\Phi_\mathrm{ext}=\Phi_0/2$. The even parity of both states results in the vanishing charge matrix element.}
\end{figure*}

In a superconducting circuit, the flux operator $\hat{\Phi}$ and charge operator $\hat{Q}$ play the roles of the conjugate quadratures. A cross-coupling between them is therefore desirable, yet it is not naturally accessible~\cite{https://doi.org/10.1002/cta.2359}. Indeed, when two circuits are capacitively or inductively connected (Fig.~\ref{fig:fig1}a), voltage or current oscillations induce the interactions, leading to charge-charge~\cite{PhysRevLett.107.080502, PhysRevLett.123.210501,PhysRevX.11.021026,PhysRevLett.129.010502} or flux-flux coupling terms~\cite{PhysRevLett.113.220502, PRXQuantum.5.020326, PhysRevLett.132.060602, PRXQuantum.3.037001}. Accessing this interaction, therefore, requires a single circuit element whose energy depends on both flux and charge, unlike the capacitor and inductor, whose constitutive relations involve only one variable each. While adding the remaining element of the superconducting toolbox---the standard Josephson junction---does not help, its voltage-tunable, semiconductor-based variant offers a natural platform to induce this interaction, as shown in Ref.~\cite{leroux2022}. 

In such semiconductor-based junctions, the occupation of the supercurrent-carrying Andreev states, and thus the energy of the junction $E_J$, can be tuned by an applied gate voltage $V_G(t)$~\cite{doi:10.1126/science.1113523,PhysRevLett.115.127001,PhysRevLett.115.127002,10.1063/5.0024124,d68y-sqzm}. For example, in the low-transmission regime, the energy of the junction is $\hat{U}_\mathrm{JJ}(\hat\Phi,t) = -E_J\left(V_G(t)\right)\cos(2\pi \, \hat\Phi/\Phi_0)$, where $\Phi_0$ is the flux quantum, and $\hat\Phi$ is the generalized flux operator that is defined as the time-integral of the voltage across the element. When the gate voltage is harmonically modulated around a bias value $V_G^0$ with frequency $f$ and small modulation amplitude $\delta V$, such that $V_G(t) = V_G^0 + \delta V\cos(2\pi f t)$, the energy of the junction can be expanded as $\hat{U}_\mathrm{JJ}(t)\approx \hat{U}_0 + \hat{U}_\mathrm{FC}(t)$. Here, the first term $\hat{U}_0=-E_J(V_G^0)\cos(2\pi\hat\Phi/\Phi_0)$ is the usual static Josephson energy providing the phase-periodic potential of gate-based circuits, such as gatemons and gatemoniums~\cite{PhysRevLett.115.127001,PhysRevLett.115.127002,PRXQuantum.6.010326,PhysRevApplied.14.064038}. The second term reveals the flux-charge interaction, as shown in Ref.~\cite{leroux2022}, which we can write as 
\begin{equation}
    \hat{U}_\mathrm{FC}(t) = \hbar g_\mathrm{FC} \left[\frac{\delta Q}{2e}\cdot\cos\left(\frac{2\pi\hat\Phi}{\Phi_0}\right)\right] \cos(2\pi f t),
\end{equation}
where the amplitude of the oscillating charge on the gate line is $\delta Q=C_G\cdot \delta V$ with $C_G$ the capacitance between the gate line and the junction, $g_\mathrm{FC} = -(2e/\hbar C_G)\cdot\partial E_J/\partial V_G $ the coupling constant that scales with the sensitivity of the junction energy to the gate voltage, and $\hbar$ the reduced Planck's constant (Fig.~\ref{fig:fig1}a). Such interaction in conventional superconducting quantum circuits is natively unavailable because capacitive or inductive couplings can separately connect either voltage or current oscillations but not their combination. Thus, voltage-tunable junctions provide a unique resource to realize cross-quadrature couplings.

\subsection*{Flux-charge interaction in fluxonium}

Central to demonstrating the flux-charge interaction is to isolate it from the always-present intra-quadrature capacitive couplings. We achieve this by relying on parity selection rules between odd- and even-parity eigenstates of a fluxonium circuit. In a fluxonium, a capacitor and a superinductor shunt a Josephson junction, which, in our case, has a gate-tunable InAs semiconducting barrier (Fig.~\ref{fig:fig1}b and c). For a fixed gate voltage and external flux, the static Hamiltonian of the circuit reads
\begin{equation}
    \hat{H}_0=4E_C\hat{n}^2-E_J^0\cos\hat\phi
+ \frac{1}{2}E_L\left(\hat\phi -2\pi \frac{\Phi_\mathrm{ext}}{\Phi_0}\right)^2,
\end{equation}
where $\hat{n}=\hat{Q}/2e$ and $\hat\phi=2\pi\,\hat{\Phi}/\Phi_0$ are the conjugate Cooper pair number and phase operators across the junction with $[\hat\phi,\hat{n}]=\mathrm{i}$,  $\Phi_\mathrm{ext}$ is the external flux piercing the loop formed by the inductor and the junction, $E_J^0=E_J(V_G^0)$ is the Josephson energy at the gate voltage, and $E_C$ and $E_L$ are the capacitive and inductive energies. At integer and half-integer flux bias, the fluxonium Hamiltonian is invariant under a unitary reflection about the corresponding symmetry point of the potential. Under this transformation, the capacitive drive operator $\hat n$ is odd, whereas the flux-charge drive operator $\cos\hat\phi$ is even. As a result, the allowed transitions induced through purely capacitive or flux-charge coupling differ, enabling us to isolate the flux-charge interaction. 

\begin{figure*}
\centering
\includegraphics[width = \textwidth]{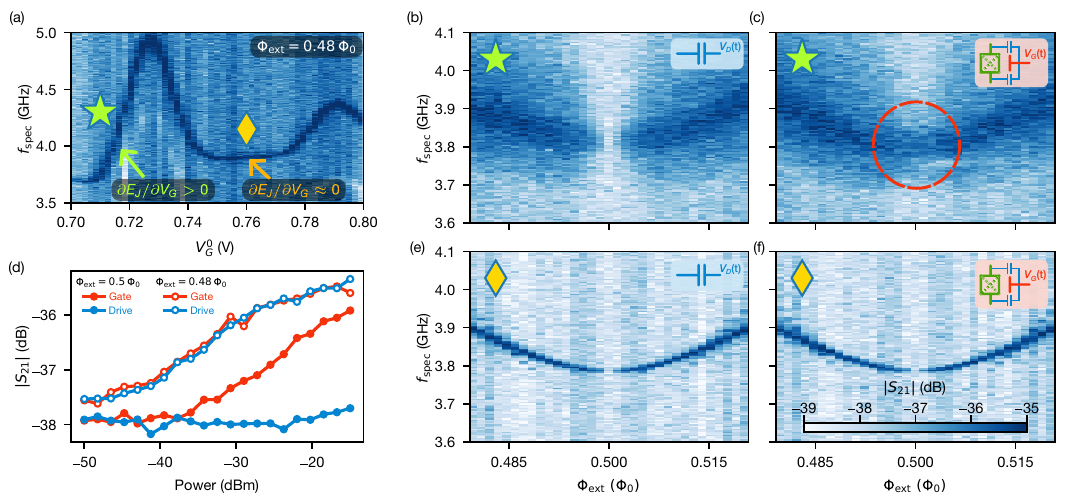}
\caption{
\label{fig:fig2} 
\textbf{Isolating flux-charge interaction in the fluxonium} a, Two-tone spectroscopy of the second excited state as a function of gate voltage at $\Phi_\mathrm{ext}\approx 0.48\,\Phi_0$. Two working points with the same Josephson energy are identified at a high-slope region (green star) and a flat region (yellow diamond). b, e, Two-tone spectroscopy as a function of external flux around half-flux quantum measured by purely capacitive excitation (through the drive line), at the high-slope and flat working points. The $|0\rangle\rightarrow|2\rangle$ transition exactly at half-flux quantum is forbidden. c, f, Two-tone spectroscopy as a function of external flux when the excitation is applied through the gate line.  The $|0\rangle\rightarrow|2\rangle$ transition at frustration is detected (red dashed circle) when $\partial E_J / \partial V_G$ is high (green star), demonstrating the pure flux-charge interaction.  d, The strength of the spectroscopy signal exactly at and slightly away from frustration as a function of drive power through the gate and drive lines at the high-slope region. Even at higher drive power, the transition at frustration is prohibited through purely capacitive coupling due to parity selection rules. Away from frustration, the strength of the spectroscopy signals through the two coupling mechanisms is similar.}
\end{figure*}

To implement and distinguish between the two driving mechanisms that couple to the $\hat n$ and $\cos\hat\phi$ operators, we have added two separate control lines to our device: a standard drive line and a gate line with voltages $V_D(t)$ and $V_G(t)$, respectively (Fig.~\ref{fig:fig1}b and c). The drive line produces a conventional charge drive through its capacitive coupling to the fluxonium islands. In contrast, the gate line modulates the junction energy $E_J\left(V_G(t)\right)$, producing a flux-charge drive through $\cos\hat\phi$, while also inducing a residual charge drive through stray capacitive coupling. For small voltage modulations around the operating point, the two corresponding driving Hamiltonians generated by the drive and gate lines are
\begin{equation}
\begin{split}
\hat{H}_D(t) &=
2e\beta_D\hat{n}\, \delta V_D\cos(2\pi ft), \\
\hat{H}_G(t) &=
2e\beta_G\hat{n}\, \delta V_G\cos(2\pi ft)
- \frac{\partial E_J}{\partial V_G}\cos\hat{\phi}\,
\delta V_G\cos(2\pi ft),
\end{split}
\end{equation}
where $\delta V_D$ ($\delta V_G$) is the modulation amplitude, and $\beta_D$ ($\beta_G$) describes the ratio of the voltage amplitude on the fluxonium capacitor pads and the drive (gate) voltage (see Supplementary Information for details).

Our study mainly focuses on the transition between the ground state $|0\rangle$ and the second excited state $|2\rangle$ of the fluxonium in the vicinity of a half-flux quantum external field (Fig.~\ref{fig:fig1}d). While at a general flux value, these states are not eigenstates of the parity operator, at half-integer flux quantum values, both of them have even parity (Fig.~\ref{fig:fig1}f and g). At these special external flux values,  $\langle 2|\hat{n}|0\rangle=0$, because the charge operator $\hat{n}$ has odd parity. Consequently, capacitive driving cannot induce transitions between these states. In contrast, even at these flux values, the matrix element $\langle 2|\cos\hat{\phi}|0\rangle\neq0$, because the even parity of $\cos\hat\phi$ allows the coupling of the two even-parity states. In Fig.~\ref{fig:fig1}e, we plot these matrix elements as a function of external flux, which shows that sharply at half-integer flux the charge-charge drive is symmetry-forbidden, whereas flux-charge drive is allowed.

To experimentally demonstrate the flux-charge interaction using a continuous-wave measurement, we show that we can induce a parity-preserving transition between the $|0\rangle$ and $|2\rangle$ states exactly at half a flux quantum. This experiment is based on standard two-tone spectroscopy, where we monitor the transmission of the readout resonator $S_{21}$, while sweeping a second spectroscopy tone $f_\mathrm{spec}$. When the spectroscopic tone is on resonance with an allowed transition, we can detect a change in the transmission amplitude due to a dispersive shift between the qubit and the resonator~\cite{RevModPhys.93.025005}. As the initial step, we carry out a spectroscopy measurement as a function of gate bias $V_G^0$ slightly away from the parity-symmetry point at $\Phi_\mathrm{ext}\approx 0.48\,\Phi_0$ (Fig.~\ref{fig:fig2}a). At this flux value, the $|0\rangle\rightarrow|2\rangle$ transition is allowed through both capacitive and flux-charge interactions because the states lack parity symmetry. Accordingly, we can measure the gate-voltage-dependence of this transition, which shows a typical non-monotonic behavior due to the filling of mesoscopic channels in the nanowire~\cite{Goffman_2017}. On this data, we highlight two regions of interest: a highly-sloped one (green star) and a flat one (yellow diamond). Since the flux-charge coupling is proportional to the gate-voltage-sensitivity of the junction, $g_\mathrm{FC}\propto \partial E_J/ \partial V_G$, and the transition energy at this flux value is proportional to the junction energy, we expect a much stronger coupling in the highly-sloped case than in the flat region.

\begin{figure*}
\centering
\includegraphics[width = \textwidth]{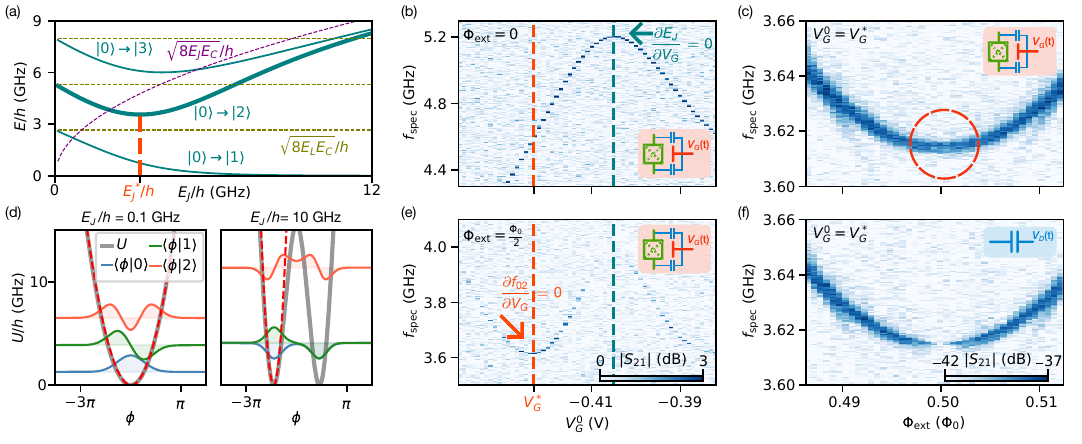}
\caption{
\label{fig:fig3} 
\textbf{The ChIVE transition in fluxonium} a, Fluxonium transitions (cyan) as a function of Josephson energy at $\Phi_\mathrm{ext}=\Phi_0/2$. The asymptotic transition frequencies are shown as gold and purple dashed lines, corresponding to the low and high Josephson energy limit, respectively. As $E_J$ increases, the fluxonium transitions move between these asymptotic values. The Charge-Insensitive-with-Variable-$E_J$ (ChIVE) point at $E^*_J$ is identified as a local minimum in the $|0\rangle\rightarrow|2\rangle$ transition frequency, where $\partial f_{02} / \partial E_J=0$ (red dashed line). d, Fluxonium potential and wavefunctions in the low (left) and high Josephson energy regime (right). The potential shifts from being approximated by a single harmonic well with curvature determined by $E_L$ to having a double-well structure with curvature determined by $E_J$. Red dashed lines show the quadratic approximation of the wells. In these asymptotic cases, the transition frequency of the $|0\rangle\rightarrow|2\rangle$ transition shifts from $\sqrt{32E_CE_L}$ to $\sqrt{8E_CE_J}$. b, Spectroscopy scan of the $|0\rangle\rightarrow|1\rangle$ transition as a function of gate voltage at $\Phi_\mathrm{ext}=0$. The frequency of this transition scales with the junction energy $f_{01}\approx\sqrt{8E_JE_C}$, hence it directly maps the gate-dependence of $E_J$. The cyan dashed line indicates where $\partial f_{01} / \partial V_G=0$, while the red dashed line shows the ChIVE point, where $\partial f_{01} / \partial V_G\neq0$ but $\partial f_{02} / \partial E_J=0$.  e, Spectroscopy scan as a function of gate voltage at $\Phi_\mathrm{ext}=\Phi_0/2$ probing the $|0\rangle\rightarrow|2\rangle$ transition through the gate line. At the ChIVE point ($V_G^0=V_G^*$, red dashed line), the transition is detectable and has a local minimum. In contrast, the signal disappears at the gate voltage where the flux-charge interaction is deactivated due to the vanishing $\partial E_J / \partial V_G$ value (cyan dashed line). c, f, Two-tone spectroscopy versus external flux on the $|0\rangle\rightarrow|2\rangle$ transition driven through the gate and drive lines at $V_G^*$, respectively. In agreement with the parity selection rules, the transition at $\Phi_\mathrm{ext}=\Phi_0/2$ is present only when the circuit is excited through the gate line, when the flux-charge interaction is enabled (red circle).}
\end{figure*}

Using these highly-sloped vs.~flat working points and the capability to address the transitions through the gate and drive lines, we perform several experiments to unambiguously demonstrate the presence of flux-charge coupling. First, we measure the response of the circuit to purely capacitive excitation $V_D(t)$ (Fig.~\ref{fig:fig2}b and e). In this case, we observe that the $|0\rangle\rightarrow|2\rangle$ transition sharply disappears at frustration because the capacitive interaction does not couple states with the same parity. This behavior is apparent both in the flat region ($\partial E_J/ \partial V_G\approx 0$) and in the high slope region ($\partial E_J/ \partial V_G > 0)$. Next, we excite the circuit through the gate line voltage $V_G(t)$, which activates both capacitive and flux-charge interactions (Fig.~\ref{fig:fig2}c and f). At the highly-sloped working point, where $\partial E_J/ \partial V_G$ is large, remarkably, we can detect a signal even at exactly half a flux quantum (red circle in Fig.~\ref{fig:fig2}c). This is the signature of transition between states of the same parity through solely flux-charge coupling. In contrast, at the other working point, where the junction energy depends less on the gate voltage ($\partial E_J/ \partial V_G\approx 0$), the spectroscopy signal at frustration fades out because the flux-charge interaction is suppressed (Fig.~\ref{fig:fig2}f). In this case, the transition away from the symmetry point is due to the parasitic capacitive coupling of the gate line and the pads of the fluxonium. Finally, we map the power-dependence of the spectroscopy signal for the case when $\partial E_J/ \partial V_G$ is large by exciting the transition through the gate and the drive lines (Fig.~\ref{fig:fig2}d). While away from the symmetry point, the two cases show similar power-dependence, at frustration, the signal increases rapidly only for the gate-line excitation due to flux-charge coupling.

\subsection*{Flux-charge interaction and charge-insensitivity}

Because the flux-charge interaction depends linearly on the voltage-sensitivity of the junction, $g_\mathrm{FC} \propto \partial E_J/\partial V_G $, operating the circuit at large coupling strengths could expose the transition to strong gate noise. Here, we show that it is possible to realize flux-charge coupling that scales linearly with the voltage drive amplitude while keeping the transition first-order protected against gate noise. The explanation behind this seemingly contradictory statement is that the decoherence rate $\Gamma_\phi$ stems from the voltage-sensitivity of the transition frequency $f_q$, while the flux-charge coupling rate results from the voltage-sensitivity of the junction energy $E_J$. To first order, these two sensitivities are connected through the chain rule
\begin{equation}
    \Gamma_\phi \propto \frac{\partial f_q}{\partial V_G}= \frac{\partial f_q}{\partial E_J}\cdot \frac{\partial E_J}{\partial V_G}\propto \frac{\partial f_q}{\partial E_J}\cdot g_\mathrm{FC}.
\end{equation}
Hence, the decoherence and the coupling rates are related through the sensitivity of the qubit frequency to the junction energy $\partial f_q / \partial E_J$. Consequently, as long as we operate the circuit at a point where $\partial f_q / \partial E_J=0$, we can ensure first-order protection against charge-noise dephasing, even when the junction has strong voltage sensitivity. While it is not possible to have a working point where $\partial f_q / \partial E_J=0$ in a gatemon, this requirement can be fulfilled in a gate-tunable fluxonium.

To understand the appearance of the $\partial f_q/\partial E_J$ sweet spot in the fluxonium, we consider how the energy levels at $\Phi_\mathrm{ext}=\Phi_0/2$ evolve as a function of the junction energy (Fig.~\ref{fig:fig3}a). In the small Josephson energy regime ($E_J\ll E_C, E_L$), the potential of the circuit can be approximated by a single harmonic well (Fig.~\ref{fig:fig3}d, left plot) where the transition frequencies are close to $\sqrt{8E_CE_L}$. In the opposite, large Josephson energy regime ($E_J\gg E_C, E_L$), the potential energy has a double-well structure (Fig.~\ref{fig:fig3}d, right plot), where the energy eigenstates come in symmetric and antisymmetric pairs, and the intra-well excitations scale with $\sqrt{8E_CE_J}$. For example, the $|0\rangle\rightarrow|2\rangle$ transition as a function of $E_J$ changes from a frequency of $\sqrt{32E_CE_L}$ to an asymptotic value of $\sqrt{8E_CE_J}$. These two limiting cases eventually lead to the emergence of a local minimum at a value labeled $E_J^*$, where $\partial f_{02} / \partial E_J=0$. At this point, which we refer to as the Charge-Insensitive-with-Variable-$E_J$ (ChIVE) point, the transition is first-order protected against gate-voltage noise, while remaining first-order sensitive to flux-charge drive.

To experimentally demonstrate the existence of the ChIVE point, we excite the circuit through the gate line. First, we directly map the bias voltage dependence of the junction energy $E_J(V_G^0)$ by measuring the $|0\rangle\rightarrow|1\rangle$ transition as a function of gate voltage at $\Phi_\mathrm{ext}=0$ (Fig.~\ref{fig:fig3}b). Here, the qubit frequency approximately scales with the junction energy as $f_{01}(\Phi_\mathrm{ext}=0)\propto \sqrt{E_J}$, allowing us to understand the gate-dependence of the junction energy.  The spectroscopy data displays a strongly gate-voltage-sensitive region (around the red dashed line, where $\partial E_J / \partial V_G \approx 76\,$GHz/V), and a charge-insensitive point (at the green dashed line, where $\partial E_J / \partial V_G = 0$). Then, we immediately measure in the same gate-voltage window the $|0\rangle\rightarrow|2\rangle$ transition at $\Phi_\mathrm{ext}=\Phi_0/2$ (Fig.~\ref{fig:fig3}e). This $f_{02}(\Phi_\mathrm{ext}=\Phi_0/2)$ transition shows a strikingly different voltage dependence from the $f_{01}(\Phi_\mathrm{ext}=0)$ transition. For example, we observe a local minimum in $f_{02}(\Phi_\mathrm{ext}=\Phi_0/2)$ at a voltage bias $V_G^*$, where, in contrast, the $f_{01}(\Phi_\mathrm{ext}=0)$ transition has a strong sensitivity to the gate voltage (red dashed lines). This is the ChIVE point, where the local minimum arises not because the junction energy is insensitive to the gate voltage ($\partial E_J / \partial V_G \neq 0$), but because the transition is insensitive to the junction energy ($\partial f_{02} / \partial E_J = 0$). 

Less surprisingly, there is also a charge-insensitive maximum in the $f_{02}(\Phi_\mathrm{ext}=\Phi_0/2)$ spectrum at a gate voltage corresponding to a maximum in the $f_{01}(\Phi_\mathrm{ext}=0)$ transition (green dashed lines). Given that the voltage-dependence of $f_{01}(\Phi_\mathrm{ext}=0)$ directly maps the voltage-dependence of the junction itself, $f_{01}(\Phi_\mathrm{ext}=0)\propto\sqrt{E_J}$, this maximum arises from the voltage-insensitivity of the junction at this gate voltage ($\partial E_J / \partial V_G = 0$). This voltage-insensitivity of the junction suppresses the flux-charge coupling because $g_\mathrm{FC}\propto \partial E_J/ \partial V_G$, which is supported by the disappearance of the spectral signal at the maximum of the $|0\rangle\rightarrow|2\rangle$ transition frequency at frustration (green dashed line in Fig.~\ref{fig:fig3}e). Finally, we measure the $|0\rangle\rightarrow|2\rangle$ transition at the ChIVE point through the gate line and drive line as a function of flux (Fig.~\ref{fig:fig3}c and f). Similar to the measurements at a high-sloped region in Figs.~\ref{fig:fig2}b and c, we observe that the flux-charge signal appears exactly at half flux quantum only when the circuit is driven through the gate line (red circle in Fig.~\ref{fig:fig3}c).

\subsection*{Coherent control through flux-charge interaction}

\begin{figure}
\centering
\includegraphics[width = \columnwidth]{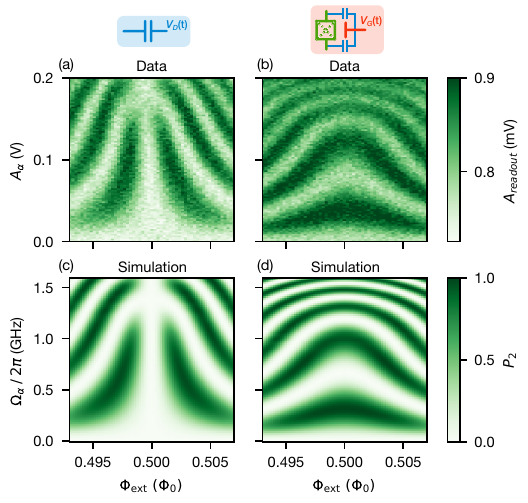}
\caption{
\label{fig:fig4} 
\textbf{Coherent control with flux-charge coupling on the $|0\rangle\rightarrow|2\rangle$ transition} a, b, Amplitude Rabi measurement as a function of external flux on the $|0\rangle\rightarrow|2\rangle$ transition at $V_G^*$ by applying the excitation through the gate and drive lines. The drive frequency of the Rabi pulse is kept fixed throughout the flux sweep at the transition frequency measured at frustration. Exactly at the half-flux point, coherent Rabi oscillations appear through flux-charge coupling but disappear when purely capacitive coupling is used to excite the transition.  c, d, Numerical simulation of the Rabi experiments by applying (c) only capacitive or (d) both capacitive and flux-charge drive terms. In the simulations, we consider purely Hamiltonian dynamics, where eight energy levels of the fluxonium are used to account for Stark-shifted levels.}
\end{figure}

Having confirmed the existence of the ChIVE point, we now demonstrate that the flux-charge interaction can be used to control a transition coherently. To carry out time-domain Rabi measurements, we voltage bias the junction at the ChIVE point $V_G^*$ to ensure that the $|0\rangle\rightarrow|2\rangle$ transition has first-order protection against gate-voltage noise.  We then use a 48\,ns long Gaussian pulse with carrier frequency $f_d$ that is resonant with the $|0\rangle\rightarrow|2\rangle$ transition at half flux quantum, such that $f_d=f_{02}(\Phi_\mathrm{ext}=\Phi_0/2)=3.61\,$GHz. Then, we map the population change of the ground state as a function of the external flux and the amplitude of the pulse while keeping the frequency of the pulse constant at the value of $f_d=3.61\,$GHz. We carry out identical flux vs.~drive-amplitude experiments by exciting the circuit through the gate and the drive lines (Fig.~\ref{fig:fig4}a and b). At a flux offset of around 5\,m{$\Phi_0$} from the symmetry point, when the detuning between the frequencies of the pulse and the transition reaches a few MHz, the two experiments show similar patterns. As the frequency detuning is increased, the oscillations become faster as expected for an off-resonant Rabi measurement.  However, the two behaviors become significantly different upon approaching $\Phi_\mathrm{ext}=\Phi_0/2$. In the case of purely capacitive driving (Fig.~\ref{fig:fig4}a), the Rabi oscillation disappears exactly at frustration because capacitive coupling cannot induce a transition between the same parity states. In contrast, when driving through the gate line (Fig.~\ref{fig:fig4}b), the oscillation remains finite at frustration, due to the transition induced by pure flux-charge interaction. Figures \ref{fig:fig4}c and d show the result of the numerical simulation of the time evolution of the system, which reveals a matching pattern to the measurements (see Methods for details). These experiments demonstrate that pure flux-charge-coupled driving can be harnessed to coherently control a transition in the presence of first-order voltage-bias insensitivity.

\subsection*{Conclusions}

In this work, we demonstrated the coherent interaction between a classical charge degree of freedom and a quantum flux operator using a voltage-tunable Josephson junction as the medium for the coupling. This implementation expands the toolbox of superconducting quantum circuits beyond the traditional same-quadrature couplings, such as charge-charge or flux-flux interactions. An essential component of the achieved unconventional coupling is that it can be activated \textit{in-situ} at a first-order gate-charge-insensitive point, allowing us to reduce the strong critical current noise susceptibility of superconductor-semiconductor hybrid devices. Already in this classical-drive regime, the coupling enables several innovations: it can facilitate interactions between fixed-frequency modes, grant symmetry-selective control of otherwise dark transitions, and govern the switching between transverse and longitudinal readout methods. When the charge oscillation is instead induced by the quantum fluctuations on the capacitor pads of a transmon, the coupling becomes fully quantum, with an estimated rate of 100\,kHz to 1\,MHz for typical device parameters~\cite{leroux2022}. Realizing the interaction in this form would take the coupling from the parametric drive demonstrated here to a native interaction between two quantum operators, completing the conjugate-variable coupling that the superconducting toolbox has so far lacked. Such a primitive could, in turn, advance alternative squeezing techniques, bosonic state control, non-reciprocal devices, and synthetic-gauge circuit elements.

\subsection*{Acknowledgements}

We thank Charles Marcus, Catherine Leroux, Ross Shillito, and Agustin Di Paolo for inspiring conversations. We gratefully acknowledge support from the U.S. Army Research Office Grant No.~W911NF-22-1-0042, the NSF Faculty Early Career Development (CAREER) Program under Award Number 2440002, the U.S. Department of Education Graduate Assistance in Areas of National Need (GAANN) grant, the Novo Nordisk Foundation (Grant No.~NNF22SA0081175, the NNF Quantum Computing Programme, NQCP), Villum Foundation through a Villum Young Investigator grant (Grant No.~37467), the Innovation Fund Denmark (Grant No.~2081-00013B, DanQ), the European Union through an ERC Starting Grant (Grant No.~101077479, NovADePro), the Carlsberg Foundation (Grant No.~CF21-0343) and the Ministère de l’Économie et de l’Innovation du Québec. A. P.-R. acknowledges support from the European Union’s Marie Skłodowska-Curie Actions (MSCA) under grant agreement No.~101204967 (FTMcQED). The views and conclusions contained in this document are those of the authors and should not be interpreted as representing the official policies, either expressed or implied, of the Army Research Office or the U.S. Government. The U.S. Government is authorized to reproduce and distribute reprints for Government purposes, notwithstanding any copyright notation herein. Views and opinions expressed are those of the author(s) only and do not necessarily reflect those of the European Union or the European Research Council. Neither the European Union nor the granting authority can be held responsible for them.

\subsection*{Methods}

\subsubsection{Fabrication}

The device fabrication starts with cleaning a high-resistivity ($>$20\,k$\Omega\cdot$cm) silicon wafer with a thickness of 525$\pm$25\,$\mu$m in a piranha solution (2:1 mixture of H$_{2}$SO$_{4}$:H$_{2}$O$_{2}$) for 10 minutes. This step is followed by two successive rinses in Milli-Q water for 60 seconds each, 1$\%$ HF dip for 30 seconds, and another two rinses in Milli-Q water for 60 seconds each. Then the wafer is blow-dried with N$_{2}$ gas and loaded into a Plassys MEB550S for aluminum deposition. The chamber is pumped for 15 hours, and 200\,nm aluminum is deposited at a rate of 0.20 nm/sec. To define alignment markers, PMMA A4 e-beam resist is spun at 4000\,rpm for 1 minute and baked at 185$^\circ$C for 2 minutes. The lithographic patterns are defined by an Elionix 125 kV electron-beam system. After exposure, the resist is developed in a 1:3 MIBK:IPA solution for 60 seconds, placed in IPA for 10 seconds, and plasma ashed for 60 seconds. The wafer is then loaded into an AJA electron-gun evaporation system for depositing 5\,nm titanium sticking layer and then 50\,nm gold layer. This is followed by a liftoff process using NMP, then acetone and IPA. The wafer is then diced into 10 $\times$ 10 mm$^2$ chips.

After dicing, bilayer PMMA A4 is spun on the chip at 4000 rpm for 1 minute, followed by a 185$^\circ$C bake for 2 minutes. The control layer is patterned by an Elionix 125 kV electron-beam system, followed by a development step in MIBK:IPA for 1 minute, IPA for 10 seconds, and plasma ashing for 2 minutes. The chip is wet-etched using Transene Aluminum Etchant D for 110 seconds, followed by rinses in Milli-Q water for 20 seconds at 50$^\circ$C and then rinses in Milli-Q water for 40 seconds at room temperature. After etching, the chip is cleaned with NMP at 80$^\circ$C for 2 hours, followed by cleaning in acetone and IPA. 

In the next layer, bilayer PMMA A4 resist is spun at 4000 rpm for 1 minute, and baked at 185$^\circ$C for 2 minutes. The gate layer is patterned and developed following the same steps as for the control layer. Then, the chip is loaded into the Plassys MEB550S system for depositing 150\,nm aluminum at a rate of 0.2\,nm/sec, followed by liftoff using NMP, acetone, and IPA. To pattern the dielectric layer with e-beam lithography, the same steps are followed. Then, Savannah Ultratech atomic layer deposition system is used to deposit $\sim$ 17 nm of HfO$_{2}$. 

In the next layer for the Dolan bridge junction array, 560\,nm thick MMA EL13 is spun at 5000 rpm for 70 seconds and baked at 185$^\circ$C for 2 minutes, followed by a spinning of 97\,nm thick PMMA A3 at 4000 rpm for 68 seconds and baking at 185$^\circ$C for 30 minutes. The junction layer is patterned and developed in MIBK:IPA (1 min), IPA (10 sec), and plasma ashed (2 min). Then, the chip is loaded into the Plassys MEB550S, where the first junction layer is fabricated by depositing 20\,nm aluminum at 23$^\circ$ angle, followed by a static oxidation at a pressure of 120\,mbar for 10 minutes. The second junction layer (top electrode) is fabricated by depositing 50\,nm aluminum at -23$^\circ$ angle, followed by a final oxidation step at 120\,mbar for 10 minutes. The chip is lifted off in dioxolane, followed by placing it in acetone and IPA. 

After junction deposition, an Al/InAs (30/130\,nm) full-shell nanowire is placed on the contact pads of the control layer using a micromanipulator. The chip is then plasma ashed for 1 minute, which promotes adhesion of the nanowire, baked at 185$^\circ$ for 2 minutes, followed by spin coating with PMMA A6 and baking at 185$^\circ$. The patch layer is patterned using an Elionix 125 kV electron-beam system, developed in 1:3 MIBK:IPA (1 min), IPA (10 sec), and plasma ashed (2 min). The galvanic connection between the nanowire and the contact pads is established in the Plassys MEB550S by first argon ion milling (at beam voltage 200\,V and ion current 15\,mA), then depositing 220\,nm aluminum at a rate of 0.2 nm/sec. 

As the final step, the chip is covered with PMMA A6 (at 4000 rpm for 60 seconds, baked at 185$^\circ$C for 10 minutes) to pattern an approximately 200\,nm wide window where the shell of the nanowire is etched. The chip follows the development step of MIBK:IPA (1 min), IPA (10 sec), and plasma ash (2 min). The chip is wet-etched using transene D for 9 seconds, rinsed in Milli-Q water for 20 seconds at 50$^\circ$C, and then in Milli-Q water for 40 seconds at room temperature. Finally, the resist is stripped using NMP at 80$^\circ$ for 2 hours.

\subsubsection{Numerical simulation of the driven fluxonium dynamics}
\label{sec:methods_driven_simulation}

The driven response (Fig.~\ref{fig:fig4}c and d) was simulated by solving the time-dependent Schrödinger equation for the fluxonium circuit. The static Hamiltonian was taken to be
\begin{align}
\hat{H}_0
=
4E_C \hat n^2
+
\frac{1}{2}E_L\left(\hat\phi -2\pi \frac{\Phi_\mathrm{ext}}{\Phi_0}\right)^2
-
E_J\cos\hat\phi ,
\end{align}
where $\hat\phi$ is the superconducting phase across the junction, $\hat n$ is the conjugate Cooper-pair number operator. For each value of flux, the Hamiltonian was diagonalized numerically in a truncated harmonic-oscillator basis. The drive operators were then expressed in the corresponding eigenbasis.

The time-dependent Hamiltonian used in the simulations was
\begin{align}
\hat{H}(t)
=
\hat{H}_0
+
\hat{H}_d(t),
\end{align}
with
\begin{equation}
\begin{split}
&\hat{H}_d(t)/\hbar
=\\
&s(t)\cos(2\pi f_d t+\phi_d)
\left[
(\Omega_c+r_n\Omega_g)\hat n
+
\varepsilon\Omega_g\cos\hat\phi
\right].
\end{split}
\end{equation}
Here $\Omega_c$ is the amplitude applied through the charge line, $\Omega_g$ is the amplitude applied through the gate line, $s(t)$ is the pulse envelope, $f_d$ is the frequency of the drive, and $\phi_d$ is the phase of the drive. The first term describes the capacitive coupling to the charge operator $\hat n$. The gate drive was modeled as having two contributions: a residual capacitive component, with relative strength $r_n$, and a flux-like component proportional to $\cos\hat\phi$, with relative strength $\varepsilon$. The fitted values used in the simulations were $r_n=1.3$ and $\varepsilon=0.087$, obtained primarily from the gate-driven oscillations measured at $\Phi_{\rm ext}=0.493\,\Phi_0$ and $\Phi_{\rm ext}=0.5\,\Phi_0$. The charge-line data, modeled as a purely capacitive drive, contained no additional fitting parameters and served as a reference for the gate-driven response.

The pulse envelope was Gaussian and defined as
\begin{align}
s(t)
=
\mathcal{N}
\exp\left[
-\frac{(t-T/2)^2}{2\sigma^2}
\right],
\qquad
0\leq t\leq T,
\end{align}
and was set to zero outside that interval. We used $\sigma=T/3$, and the normalization constant $\mathcal{N}$ was chosen such that $s(T/2)=1$. The drive frequency was fixed to the $|0\rangle\rightarrow|2\rangle$ transition frequency at half flux, $f_d
=f_{02}(\Phi_{\rm ext}=\Phi_0/2)$.

The system was initialized in the ground state $|0\rangle$ of the static Hamiltonian at the corresponding flux bias. We then integrated
$i\frac{d}{dt}\ket{\psi(t)}=(H(t)/\hbar)\ket{\psi(t)}$
up to the end of the pulse, and extracted the final population $P_2(T)=|\bra{2}\ket{\psi(T)}|^2$.
The time evolution was computed with \texttt{QuTiP}, using direct integration of the Schrödinger equation with \texttt{sesolve}. Relaxation and dephasing were not included in the simulations. This approximation is justified by the fact that the coherence time of the transition is long compared with the pulse duration used in the experiment.

At the half-flux point, $\Phi_{\rm ext}=\Phi_0/2$, the static Hamiltonian is symmetric under reflection about $\phi=\pi$. The eigenstates can therefore be assigned a definite parity. Under this reflection symmetry, $\hat n$ is odd, whereas $\cos\hat\phi$ is even. Consequently, the matrix element $\langle 0|\hat n|2\rangle$ vanishes at the symmetry point, while the corresponding matrix element of $\cos\hat\phi$ is not symmetry-forbidden. The simulation, therefore, separates the conventional capacitive response from the gate-induced flux-like response.

We verified convergence with respect to the harmonic-oscillator cutoff and the number of eigenstates retained in the time evolution. In the parameter range shown, including states up to $\ket{7}$ was sufficient to obtain stable values of $P_2$. Higher excited states acquire only a small direct population during the pulse, but they are needed to capture the amplitude-dependent renormalization of the low-energy dynamics through off-resonant coupling, especially at large drive amplitude.


%



\subsection*{Supplementary Information for ``Observation of coherent flux-charge interaction in a gate-tunable fluxonium''}

\subsubsection*{Device parameters}

\begin{figure}
\centering
\includegraphics[width = \columnwidth]{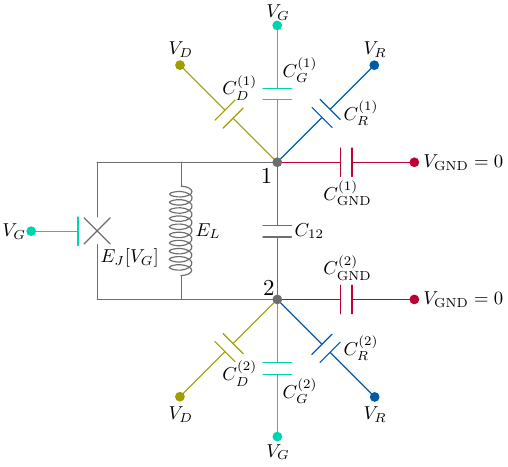}
\caption{
\label{fig:figs1} 
Circuit schematic of the gate tunable fluxonium device.}
\end{figure}

For this study, we measure a fluxonium qubit with floating capacitor pads, where a voltage-tunable Josephson junction is placed in parallel with an inductor between nodes 1 and 2 (Fig.~\ref{fig:figs1}). The junction energy is $E_J(V_G)$, which is tunable with gate voltage $V_G$, the inductive energy is $E_L$, and the capacitive energy is $E_C$. We define the generalized node flux variables ${\Phi}_1$ and ${\Phi}_2$ as the time-integral of the voltages at the nodes. A separate ground plane provides the reference voltage, $V_\mathrm{GND}=0$. The device is capacitively coupled to the center pin of a readout resonator with voltage $V_R$, to the drive line with voltage $V_D$, and to the gate line with voltage $V_G$. We define the flux and voltage vectors as $\mathbf{\Phi}=\left(\Phi_1, \Phi_2\right)^\mathrm{T}$, $\mathbf{V}_R^\Phi=(V_R, V_R)^\mathrm{T}$, $\mathbf{V}_D^\Phi=(V_D, V_D)^\mathrm{T}$, $\mathbf{V}_G^\Phi=(V_G, V_G)^\mathrm{T}$. Furthermore, $C_{12}$ is the capacitance between nodes 1 and 2, while $C_\mathrm{GND}^{(i)}$, $C_R^{(i)}$, $C_D^{(i)}$, $C_G^{(i)}$ are the capacitances between node i and the ground, the resonator, the drive line, and the gate line, respectively. Thus, the capacitance matrix of the fluxonium is
\begin{equation}
\mathbf{C}^\Phi=
    \begin{pmatrix}
     C_{11}  & -C_{12} \\
    -C_{12} & C_{22}\\
    \end{pmatrix},
\end{equation}
where $C_{11} = C_{12} + C_\mathrm{GND}^{(1)} + C_R^{(1)} + C_D^{(1)} + C_G^{(1)}$ and $C_{22}=C_{12} + C_\mathrm{GND}^{(2)} + C_R^{(2)} + C_D^{(2)} + C_G^{(2)}$.
The capacitance matrices associated with the coupling to the external voltages are $\mathbf{C}_R^\Phi=\text{diag}\left(C_R^{(1)}, C_R^{(2)}\right)$, $\mathbf{C}_D^\Phi=\text{diag}\left(C_D^{(1)}, C_D^{(2)}\right)$, $\mathbf{C}_G^\Phi=\text{diag}\left(C_G^{(1)}, C_G^{(2)}\right)$. With these, the Lagrangian of the circuit is 
\begin{equation*}
\begin{split}
    L^\Phi&=\frac{1}{2}\Dot{\mathbf{\Phi}}^\mathrm{T}\mathbf{C}^\Phi\Dot{\mathbf{\Phi}}-\Dot{\mathbf{\Phi}}^\mathrm{T}\mathbf{C}_R^\Phi\mathbf{V}_R^\Phi-\Dot{\mathbf{\Phi}}^\mathrm{T}\mathbf{C}_D^\Phi\mathbf{V}_D^\Phi-\Dot{\mathbf{\Phi}}^\mathrm{T}\mathbf{C}_G^\Phi\mathbf{V}_G^\Phi \\+ &E_J\cos\left[\frac{2\pi}{\Phi_0}\left(\Phi_1-\Phi_2\right)\right] - \frac{1}{2}E_L\left[\frac{2\pi}{\Phi_0}\left(\Phi_1-\Phi_2-\Phi_\mathrm{ext}\right)\right]^2.
\end{split}
\end{equation*} 

\begin{figure*}
\centering
\includegraphics[width = 0.77\textwidth]{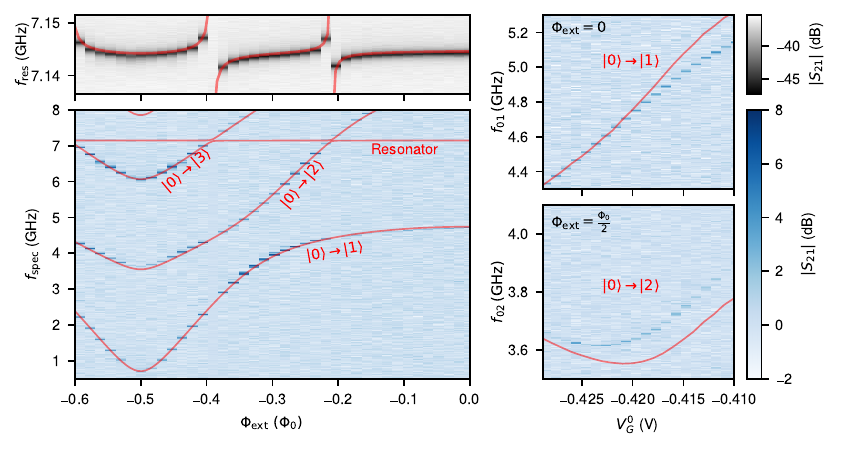}
\caption{
\label{fig:figs2} 
(Left) Energy levels of the circuit as a function of external flux at the ChIVE point. The top panel shows the flux-dependence of the resonator frequency, while the bottom panel displays the background-subtracted result of a two-tone spectroscopy. Red curves are theory curves obtained from fitting the two-tone spectroscopy data on the fluxonium energy levels using scQubits~\cite{Groszkowski2021scqubitspython}. (Right) Fit of the spectrum as a function of gate voltage around the ChIVE point (same data as Fig.~\ref{fig:fig3}b and e in the main text). For this fit, only one free parameter ($E_J$) is used, while the other circuit parameters are obtained from the flux-dependent spectrum. The slight deviation between theory and experiment may arise from higher harmonic contributions of the nanowires, nonlinearities in the Josephson junction array, or small gate voltage drifts.}
\end{figure*}

Next, we introduce the branch flux $\Phi_\Delta=\Phi_1-\Phi_2$, and the center of mass flux $\Phi_\Sigma=(\eta_1\Phi_1+\eta_2\Phi_2) / (\eta_1+\eta_2)$, where $\eta_i=(C_\mathrm{GND}^{(i)} + C_R^{(i)} + C_D^{(i)} + C_G^{(i)})$. Their vector is $\mathbf{\Theta}=\left(\Phi_\Delta, \Phi_\Sigma\right)^\mathrm{T}$. This change of basis is described by the transformation $\mathbf{\Theta}=\mathbf{R}\cdot\mathbf{\Phi}$, where 
\begin{equation*}
    \mathbf{R}=
    \begin{pmatrix}
    1 & -1 \\
    \eta_1/(\eta_1+\eta_2) & \eta_2/(\eta_1+\eta_2)
    \end{pmatrix}.
\end{equation*}
The Lagrangian in this basis reads
\begin{equation*}
\begin{split}
    L^\Theta&=\frac{1}{2}\Dot{\mathbf{\Theta}}^\mathrm{T}\mathbf{C}^\Theta\Dot{\mathbf{\Theta}}-\Dot{\mathbf{\Theta}}^\mathrm{T}\mathbf{C}_R^\Theta\mathbf{V}_R^\Theta-\Dot{\mathbf{\Theta}}^\mathrm{T}\mathbf{C}_D^\Theta\mathbf{V}_D^\Theta-\Dot{\mathbf{\Theta}}^\mathrm{T}\mathbf{C}_G^\Theta\mathbf{V}_G^\Theta \\& +E_J\cos\left(\frac{2\pi}{\Phi_0}\Phi_\Delta\right) - \frac{1}{2}E_L\left[\frac{2\pi}{\Phi_0}\left(\Phi_\Delta-\Phi_\mathrm{ext}\right)\right]^2.
\end{split}
\end{equation*} 
Here, $\mathbf{C}^\Theta = (\mathbf{R}^{-1})^\mathrm{T}\mathbf{C}^\Phi\mathbf{R}^{-1}$, $\mathbf{C}_R^\Theta = (\mathbf{R}^{-1})^\mathrm{T}\mathbf{C}_R^\Phi\mathbf{R}^{-1}$, $\mathbf{C}_D^\Theta = (\mathbf{R}^{-1})^\mathrm{T}\mathbf{C}_D^\Phi\mathbf{R}^{-1}$,
$\mathbf{C}_G^\Theta = (\mathbf{R}^{-1})^\mathrm{T}\mathbf{C}_G^\Phi\mathbf{R}^{-1}$,
$\mathbf{V}_R^\Theta=\mathbf{R}\mathbf{V}_R^\Phi$, and $\mathbf{V}_D^\Theta=\mathbf{R}\mathbf{V}_D^\Phi$, and $\mathbf{V}_G^\Theta=\mathbf{R}\mathbf{V}_G^\Phi$. In this Lagrangian, the $\Phi_{\Delta}$ and $\Phi_\Sigma$ modes decouple, and the $\Phi_\Sigma$ mode does not have a potential term associated with it.

To arrive at the Hamiltonian of the circuit, we introduce the conjugate charge variables $\mathbf{q}_\Theta=(\partial L^\Theta/ \partial \dot{\Phi}_\Delta, \partial L^\Theta/ \partial \dot{\Phi}_\Sigma)^\mathrm{T} = (q_\Delta, q_\Sigma)^\mathrm{T}$ and perform a Legendre transformation to arrive at the Hamiltonian $H = H_0 + H_R + H_D + H_G$, where
\begin{equation}
\begin{split}
    H_0  &=
\frac{1}{2}\mathbf{q}_\Theta^\mathrm{T}(\mathbf{C}^\Theta)^{-1}\mathbf{q}_\Theta
- E_J(V_G)\cos\left(\frac{2\pi}{\Phi_0}\Phi_\Delta\right)\\
&+ \frac{1}{2}E_L\left[\frac{2\pi}{\Phi_0}\left(\Phi_\Delta-\Phi_\mathrm{ext}\right)\right]^2,\\   
    H_R &=\mathbf{q}_\Theta^T(\mathbf{C}^\Theta)^{-1}\mathbf{C}^\Theta_R\mathbf{V}^\Theta_R,   \\ 
    H_D &=\mathbf{q}_\Theta^T(\mathbf{C}^\Theta)^{-1}\mathbf{C}^\Theta_D\mathbf{V}^\Theta_D, \\
    H_G &=\mathbf{q}_\Theta^T(\mathbf{C}^\Theta)^{-1}\mathbf{C}^\Theta_G\mathbf{V}^\Theta_G.
\end{split}
\end{equation}

Using circuit quantization, we promote the variables into operators, introduce the Cooper pair number and phase operators $\hat{q}_\Delta=2e\hat{n}$ and $\hat{\Phi}_\Delta=\Phi_0\hat{\phi}/2\pi$, which allows us to write the Hamiltonian for the differential mode
\begin{equation}
\begin{split}
    \hat{H}_0  &= 4E_C\hat{n}^2 - E_J(V_G)\cos\hat{\phi} + \frac{1}{2}E_L\left(\hat{\phi}-\frac{2\pi}{\Phi_0}\Phi_\mathrm{ext}\right)^2, \\   
    \hat{H}_R &=2e\beta_RV_R\hat{n},   \\ 
    \hat{H}_D &=2e\beta_DV_D\hat{n}, \\
    \hat{H}_G &=2e\beta_GV_G\hat{n}.
\end{split}
\end{equation}
Here, the charging energy is $E_C=e^2/2C_\Phi$, where $C_\Phi=1/\left[(\mathbf{C}^\Theta)^{-1}\right]_{00}$, and the coupling constants are 
\begin{equation}
\begin{split}
\beta_R
&=
\frac{\eta_2 C_R^{(1)}-\eta_1 C_R^{(2)}}{(\eta_1+\eta_2)C_\Phi},
\\
\beta_D
&=
\frac{\eta_2 C_D^{(1)}- \eta_1 C_D^{(2)}}{(\eta_1+\eta_2)C_\Phi},
\\
\beta_G
&=
\frac{\eta_2 C_G^{(1)} - \eta_1 C_G^{(2)}}{(\eta_1+\eta_2)C_\Phi}.
\end{split}
\end{equation}
From finite-element capacitive matrix simulations, we find $E_C/h=0.87\,$GHz, $\beta_R=0.11$, $\beta_D=-4.2\times10^{-3}$, and $\beta_G=-1.1\times10^{-3}$. From measuring both the resonator and the fluxonium energy levels as a function of flux (Fig.~\ref{fig:figs2}), we find $E_C/h = 0.883\,$GHz, $E_L/h = 0.999\,$GHz, and $\beta_R=0.096$, in agreement with the finite-element models. 

\begin{figure*}
\centering
\includegraphics[width = \textwidth]{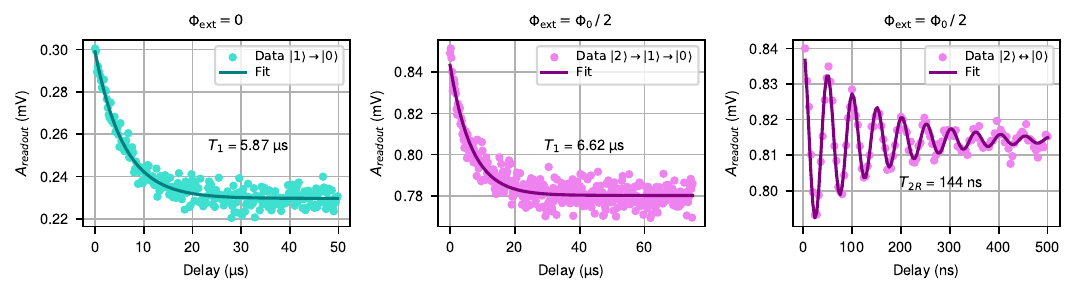}
\caption{
\label{fig:figs3} 
Measured $T_1$ value for the plasmon transition at zero flux (cyan) and the measured $T_1$ and $T_2$ Ramsey values for the $|0\rangle\rightarrow|2\rangle$ transition at the ChIVE point (purple). The measured relaxation rate for the $|0\rangle\rightarrow|2\rangle$ transition is the combination of relaxation from the $|2\rangle\rightarrow|1\rangle$ and $|1\rangle\rightarrow|0\rangle$ transitions.}
\end{figure*}

\subsubsection*{Coherence times}

In Fig.~\ref{fig:figs3}, we report measured lifetimes of the device at various operating points and transitions, illustrating that the relaxation and coherence times are comparable to the standard values in the field.

\subsubsection*{Measurement setup}

\begin{figure*}
\centering
\includegraphics[width = 0.77\textwidth]{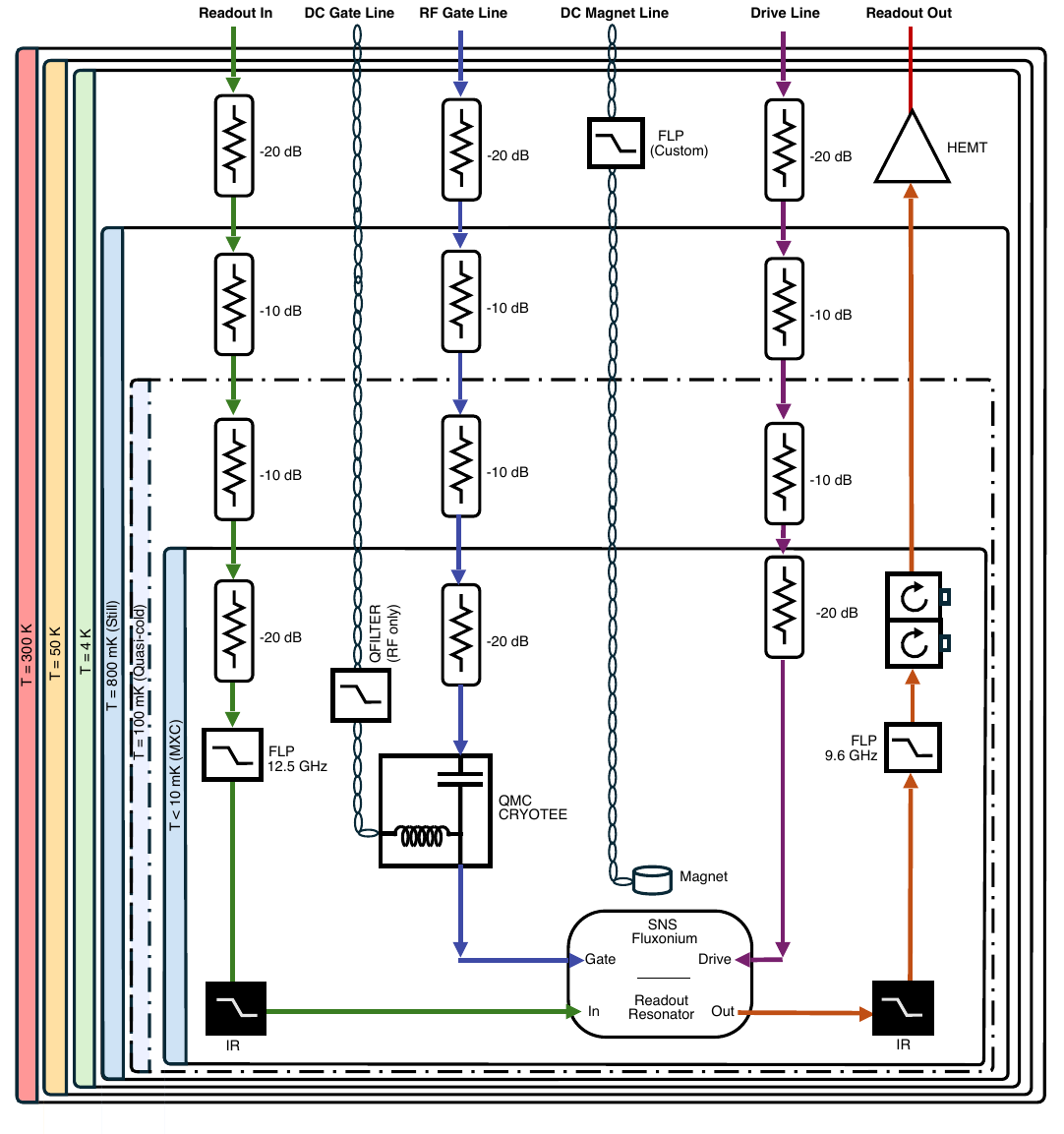}
\caption{
\label{fig:figs4} 
Schematics of the fridge wiring diagram used for the experiments reported in this work.}
\end{figure*}

Figure \ref{fig:figs4} shows the schematics of the low-temperature measurement setup in a Bluefors LD-400 dilution refrigerator. The continuous wave experiments were carried out using a Rohde \& Schwarz ZNB Vector Network Analyzer, while the time-domain measurements were obtained using the Quantum Machines OPX and Octave systems. QDevil QDAC system supplied the DC bias gate voltage and the current for the external flux. The sample was housed in a QCage sample holder with a built-in magnet. At room temperature, a Mini-Circuits RF Switch Matrix allowed for the quick switching between measurements through the gate and drive lines. 

\end{document}